\documentclass[fleqn,usenatbib]{mnras}

\usepackage{newtxtext}
\usepackage[slantedGreek]{newtxmath}

\usepackage[T1]{fontenc}
\usepackage{float}
\DeclareRobustCommand{\VAN}[3]{#2}
\let\VANthebibliography\thebibliography
\def\thebibliography{\DeclareRobustCommand{\VAN}[3]{##3}\VANthebibliography}

\usepackage{graphicx}	
\usepackage{amsmath}	
\usepackage[normalem]{ulem}

\hypersetup{pdfauthor={Y. C. Perrott},
               pdftitle={Sub-kpc radio jets in the brightest central galaxy of the cool-core galaxy cluster RXJ1720.1+2638},
               pdfkeywords={radiation mechanisms: non-thermal -- galaxies: active -- galaxies: jets -- galaxies: clusters: individual: RXJ1720.1+2638 -- radio continuum: galaxies},
               bookmarksnumbered=true}

\title[Sub-kpc radio jets in RXJ1720.1+2638]{Sub-kpc radio jets in the brightest central galaxy of the cool-core galaxy cluster RXJ1720.1+2638}

\author[Perrott et al.]{Yvette~C.~Perrott,$^{1}$\thanks{E-mail: yvette.perrott@vuw.ac.nz}
Gopika~SM,$^{2}$
Alastair~C.~Edge,$^{3}$
Keith~J.~B.~Grainge,$^{4}$
\newauthor
David~A.~Green,$^{5}$
Richard~D.~E.~Saunders$^{5,6}$
\\
$^{1}$School of Chemical and Physical Sciences, Victoria University of Wellington, PO Box 600, Wellington 6140, New Zealand\\
$^{2}$Department of Physical Sciences, Indian Institute of Science Education and Research Kolkata, Mohanpur, 741 246, West Bengal, India\\
$^{3}$Centre for Extragalactic Astronomy, Durham University, DH1 3LE, UK\\
$^{4}$Jodrell Bank Centre for Astrophysics, Department of Physics \& Astronomy, The University of Manchester, Manchester M13 9PL, UK\\
$^{5}$Astrophysics Group, Cavendish Laboratory, JJ Thomson Avenue, Cambridge CB3 0HE, UK\\
$^{6}$Kavli Institute for Cosmology Cambridge, Madingley Road, Cambridge CB3 0HA, UK
}

\date{Accepted XXX. Received YYY; in original form ZZZ}

\pubyear{2022}

\makeatletter\NAT@longnamestrue\makeatother

\begin{document}
\label{firstpage}
\pagerange{\pageref{firstpage}--\pageref{lastpage}}
\maketitle

\begin{abstract}
The cool-core galaxy cluster RXJ1720.1+2638 hosts extended radio emission near the cluster core, known as a minihalo.  The origin of this emission is still debated and one piece of the puzzle has been the question of whether the supermassive black hole in the brightest central galaxy is actively powering jets.  Here we present high-resolution e-MERLIN observations clearly indicating the presence of sub-kpc jets; this may have implications for the proposed origin of the minihalo emission, providing an ongoing source of relativistic electrons rather than a single burst sometime in the past, as previously assumed in simulations attempting to reproduce observational characteristics of minihalo-hosting systems.

\end{abstract}




\section{Introduction}
RXJ1720.1+2638 is a cool-core cluster of galaxies with a pair of cold fronts detected in X-ray suggesting a `sloshing' motion has been induced by a minor merger \citep{2001ApJ...555..205M}, likely with the smaller subcluster detected via gravitational lensing \citep{2010PASJ...62..811O} and galaxy density mapping \citep{2011ApJ...741..122O}.  The core of the cluster also hosts a radio minihalo (\citealt{2008ApJ...675L...9M}, \citealt{Giacintucci}), diffuse synchrotron emission at the cluster core, which appears to be confined by the cold fronts.

There are two mechanisms generally considered to be possible for the production of minihalo emission at the cores of galaxy clusters.  Firstly, the `re-acceleration' mechanism \citep{2002A&A...386..456G}: old relativistic electrons from past outbursts of the active galactic nucleus (AGN) in the brightest central galaxy (BCG) are re-accelerated to energy levels sufficient to produce observable synchrotron radiation by turbulence induced by the minor merger or other mechanisms.  Secondly, the `hadronic' or `secondary' mechanism: thermal protons throughout the cluster volume are accelerated to relativistic energies by supernovae, AGN and intracluster medium (ICM) shocks; these cosmic ray protons undergo interactions with the thermal proton population, producing pions which decay into a series of products including relativistic electrons (e.g.\ \citealt{2004A&A...413...17P}).

Several features of the minihalo emission in RXJ1720.1+2638 seem to support the re-acceleration mechanism.  The confinement of the minihalo emission by the cold fronts is supported by simulations assuming that aged electrons from a single, past injection of relativistic particles are distributed by sloshing turbulence \citep{2013ApJ...762...78Z}, and the steepening spectral index of the `tail' of the minihalo extending to the south is expected if the re-accelerated electrons lose energy as they travel further from the initial pool of aged electrons at the centre.  On the other hand, \citet{2021MNRAS.508.2862P} showed that the spectrum of the central minihalo followed a constant power-law of approximately $S_{\nu} \propto \nu^{-1.0}$ over a large range of frequencies from 317\,MHz up to at least 18\,GHz.  The spectrum is both slightly flatter than predicted by the \citet{2013ApJ...762...78Z} simulations and does not show the spectral break expected in the re-acceleration scenario.  \citet{2021MNRAS.508.2862P} also found that the minihalo was larger than expected at high frequency, and potentially extends out past the cold fronts.  These observations could point to better agreement with the hadronic mechanism, as demonstrated in simulations by \citep{2015ApJ...801..146Z}.

A complicating factor, however, was the lack of knowledge of the state of the BCG supermassive black hole.  The \citet{2013ApJ...762...78Z} simulations assumed a single AGN outburst in the past; the relativistic electrons were then allowed to age for some time before being re-accelerated by the turbulence induced by the minor merger.  If the BCG in fact has an AGN which is actively accreting and producing jets, the continuous injection of new relativistic electrons could explain the lack of spectral steepening and slightly flatter spectral index, while still requiring the turbulence to spread the electrons from the jets out to the observed extent of the minihalo.  Previous radio observations had only been able to put an upper limit on the size of any prospective jets ($<1.4$\,kpc; \citealt{Giacintucci}) although various optical indicators suggested AGN activity (e.g.\ \citealt{2012MNRAS.423..422L}, \citealt{2016MNRAS.461..560G}).

In this letter, we present high-resolution observations from the extended Multi-Element Radio Linked Interferometer Network (e-MERLIN; \citealt{2004SPIE.5489..332G}) which demonstrate unequivocally the presence of jets extending at least 0.8 pc from the BCG centre.  This continuous injection of relativistic electrons must be taken into account when interpreting the observations in favour of either acceleration mechanism.

Throughout the letter we use J2000.0 coordinates and use the convention that radio flux density $S$ depends on frequency $\nu$ as $S \propto \nu^{-\alpha}$ with spectral index $\alpha$. Unless stated otherwise we use $\Lambda$CDM cosmology with $\Omega_{\mathrm{m}} = 0.3$, $\Omega_{\Lambda} = 0.7$ and $H_0 = 70$\,km\,s$^{-1}$\,Mpc$^{-1}$. With this cosmology, at the redshift of the cluster ($z = 0.160$; \citealt{2011ApJ...741..122O}) 1\,arcsec corresponds to 2.76\,kpc. We use the `cubehelix' colour scheme defined by \citet{2011BASI...39..289G} for radio astronomical images.

\section{Observations}\label{S:observations}

The cluster centre was observed as a single pointing with e-MERLIN on 2020 Aug 13 -- 14 and 2022 Jun 22 -- 24 for a total of $\approx$\,28\,hours on-source under project code CY10222.  The frequency band was 4.5 -- 5\,GHz with a central frequency of 4.76\,GHz.  All antennas were present except the Lovell telescope; in the 2020 Aug observation, the Knockin antenna and most of the Mark II antenna data were flagged due to an instrumental error.

The shortest baseline present in the combined observations is $\approx$\,35\,k$\lambda$, corresponding to a largest angular scale of $\approx$\,6\,arcsec.  The minihalo, which has an angular size of at least 30\,arcsec, is therefore completely resolved out in these observations allowing us to isolate the core and jet/lobe emission.  The longest baseline is $\approx$\,3.6\,M$\lambda$, and with a Briggs robust weighting of 0.5 we achieve a synthesized beam size of $0.06 \times 0.04$\,arcsec$^2$ corresponding to a physical size of $165 \times 110$\,pc$^2$.

\section{Calibration and imaging}\label{S:calibration}

We use 3C\,286 as the primary flux calibration source, using the \citet{2017ApJS..230....7P} scale to set the flux density.  We use the standard e-MERLIN pipeline reduction\footnote{\url{https://github.com/e-merlin/eMERLIN_CASA_pipeline}}, using interleaved observations of a nearby point-like calibrator, 1722+2815, to apply phase calibration to the science field.  We also added an extra bandpass calibration using the interleaved calibrator to remove some residual amplitude errors.

The calibrated data were imaged using the \textit{tclean} task in \textsc{casa}\footnote{\url{https://casa.nrao.edu/}}.  We used Briggs weighting with a robust parameter of 0.5 which gave the best trade-off between beam size and sensitivity.  To further increase the sensitivity to the larger-scale structures of the jets, we also imaged with $uv$-tapers of 800 and 500\,k$\lambda$.

\section{Results and analysis}\label{S:results}

\begin{figure}
    \centering
\includegraphics[trim=2.5cm 11cm 4cm 6cm, clip, width=\columnwidth]{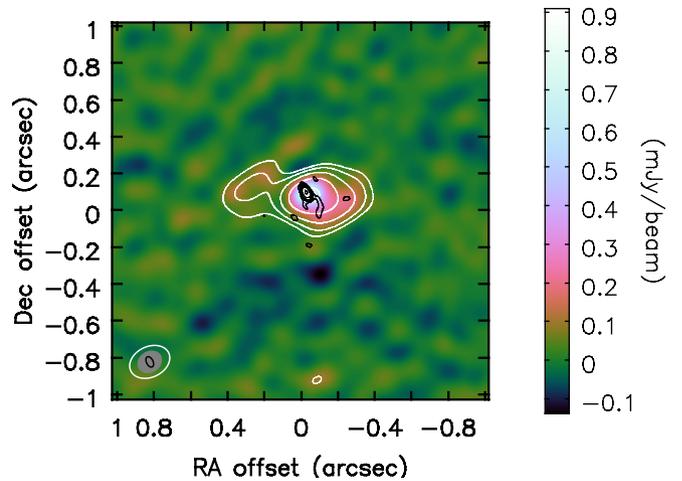}
    \caption{e-MERLIN image of the RXJ1720.1+2638 BCG at three different resolutions.  The background colour-scale shows the 800\,k$\lambda$-tapered version, showing the details of the jet structure; the black contours show the untapered image, showing the central source and the beginning of the western jet; the white contours show the 500\,k$\lambda$-tapered version, showing that there is a significant detection of the jets on both sides.  The five contours are logarithmically spaced from $3\sigma = 60.8 \: (96.6)$\,$\upmu$Jy\,beam$^{-1}$ to 90 per cent of the peak flux density, where peak flux density $= 926 \: (1008)$\,$\upmu$Jy\,beam$^{-1}$ in the untapered (500\,k$\lambda$-tapered) case.  Synthesised beams are shown in the bottom left-hand corner.}
    \label{Fi:radio_image}
\end{figure}

The resulting image is shown in Fig.~\ref{Fi:radio_image}, with colour-scale showing the 800\,k$\lambda$-tapered image and contours showing the untapered and 500\,k$\lambda$-tapered images.  There is a clear detection of a central, compact object plus resolved jets extending to the east and west.  The jets extend roughly 0.3\,arcsec (corresponding to 0.8\,kpc) from the centre in both the east and west directions.  The eastern jet may bend abruptly to the north-west, however this feature only appears at $\approx$\,2$\sigma$ significance.  In contrast, the minihalo emission extends to at least 80\,kpc from the cluster centre, so these jets are around three orders of magnitude smaller.

We fitted a Gaussian to the central point-like object in the untapered image using the \textsc{casa} task \textit{imfit}, finding a peak flux density of $961\pm 23$\,$\upmu$Jy beam$^{-1}$, with no indication of extension.  Measuring the integrated flux density of the point-like $+$ extended emission down to the $2\sigma$ contour on the 500\,k$\lambda$-taper image gave $1.76 \pm 0.08$ mJy in total (where the error is calculated by multiplying the rms noise by $\sqrt{N_\mathrm{beams}}$, $N_\mathrm{beams}$ being the number of beams covered by the source).  Subtracting the point-like from the total flux density we measure $801 \pm 90$\,$\upmu$Jy for the jets.

\begin{figure}
    \centering
    \includegraphics[trim=0cm 0cm 0cm 0cm, clip, width=\columnwidth]{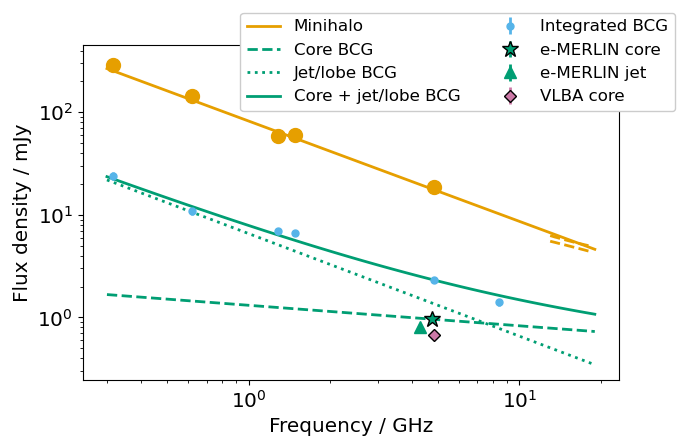}
    \caption{Spectra for the various components of the BCG system.  Large orange points and lines show the minihalo; the datapoints are measurements from \citet{Giacintucci} (and references therein), the dashed lines show the high frequency $1\sigma$ constraints from \citet{2021MNRAS.508.2862P} and the solid line is a power-law fit to the lower-frequency datapoints.  Light blue small points show the integrated central BCG component and are also from \citet{Giacintucci} and references therein.  The green star and triangle show the core and jet measurements from these e-MERLIN observations respectively, where the triangle is displaced horizontally for clarity. The dashed green and dotted green lines show spectra of the core and jet components normalised to the e-MERLIN measurements (assuming the jets are partially resolved out by e-MERLIN so that full jet $+$ core flux density adds up to the integrated measurement) and assuming representative spectral indices $\alpha_{\mathrm{core}}=0.2$ and $\alpha_{\mathrm{jet}}=1.0$.  The solid green line is the sum of these components and agrees fairly well with the integrated measurements over the whole frequency range; it is not fitted to the integrated measurements.  The purple diamond shows the VLBA core measurement, indicating some variability in comparison to the e-MERLIN measurement.}
    \label{Fi:BCG_spec}
\end{figure}

We can assess whether we have recovered all the extended flux from the jets by comparing our flux density measurement to the previous, lower-resolution flux density measurements that did not resolve the jets, as illustrated in Fig.~\ref{Fi:BCG_spec}.  Using the flux densities given in \citet{Giacintucci} to fit a power-law spectrum, we obtain a flux density of $S = 2.34 \pm 0.10$\,mJy at the e-MERLIN central frequency of 4.76\,GHz.  Subtracting the measured core, we would therefore expect a jet flux density of $2.34 - 0.961 = 1.38$\,mJy and are recovering $\approx$\,60 per cent of the jet flux in this e-MERLIN observation.  This indicates that the jets may have a larger extent than this observation detects.  The unresolved core is a significant fraction of the total, $\approx$\,40 per cent.

\citet{2014PhDT.......338H} measure a VLBA flux of $680\pm 60$\,$\upmu$Jy\,beam$^{-1}$ from a 2013 observation for the unresolved core.  This would imply variability of the core of $\sim$40 per cent over 8 years as is common for BCGs \citep{2022MNRAS.509.2869R}. From Figure 7 in \citep{2022MNRAS.509.2869R}, at this level of variability more than half of the sources show variation on $>$8 year timescales. On the other hand, the majority of these sources would show only variability of 20 per cent or less on the 2 year timescale between our e-MERLIN observations, consistent with the match in measured peak flux densities in 2020 and 2022 ($955 \pm 40$ and $924 \pm 59$\,$\upmu$Jy beam$^{-1}$ in 2020 and 2022 respectively; errors are thermal noise only). This variability impacts on our estimate of the fraction of the jet flux density recovered in our e-MERLIN observation; if the core flux density were 60 per cent higher when the integrated flux density measurement was made (in 1985) than at the e-MERLIN measurement epoch, the jet flux density measured from the e-MERLIN observation would account for the total jet flux density.

At frequencies lower than 2\,GHz, where the integrated flux density measurement is dominated by the jet/lobe component, the measurements from 1985 to 2009 show no significant variation.

\section{Discussion}\label{S:discussion}

\subsection{Host galaxy}

\begin{figure}
    \centering
    \includegraphics[trim=2cm 10cm 3.75cm 4.4cm, clip, width=\columnwidth]{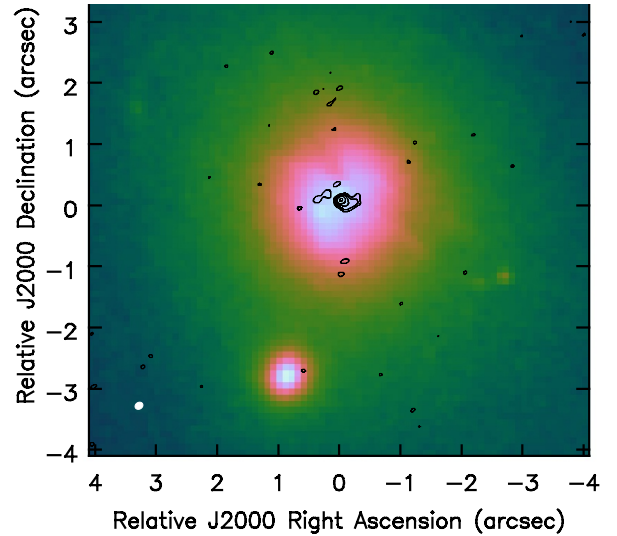}
    \caption{HST image of the BCG (colour-scale) with contours from the 800\,k$\lambda$-tapered e-MERLIN image overlaid. The five contours are logarithmically spaced from $3\sigma = 73.1$\,$\upmu$Jy\,beam$^{-1}$ to 90 per cent of the peak flux, where peak flux $=912$\,$\upmu$Jy\,beam$^{-1}$.  The synthesised beam is shown in the bottom left-hand corner as the filled white ellipse.}.
    \label{Fi:HST}
\end{figure}

A comparison to the optical image of the galaxy from the Hubble Space Telescope (HST), shown in Fig.~\ref{Fi:HST}, shows that in the current observation the jets appear to be inside the host galaxy rather than extending out into the ICM.  The AGN may however have been more active in the past, with jets and lobes extending into the ICM, and now be in a quiescent phase of the feedback cycle.  Lower-frequency, high-resolution observations may also reveal older, steep-spectrum emission from extended lobes further out in the ICM that is too faint to be seen at the relatively high frequency of these observations.  Based on the length of the jets shown in the current observations and assuming particles in the jets move at speeds $<c$, we can infer a lifetime of this phase in the AGN's duty cycle of as little as $\sim 3000$\,years.

\subsection{Correlations with minihalo power}

\citet{2020MNRAS.499.2934R} analysed correlations between the luminosity of radio minihaloes and the BCG luminosity, dividing the latter into the core component representing on-going accretion and the jet/lobe component representing past activity.  They used the technique and data from \citet{2015MNRAS.453.1201H}, spectrally separating the emission by fitting (where appropriate) a double power law and identifying the flatter component as the core and the steeper component as the non-core.  However, as noted in \citet{2015MNRAS.453.1201H}, the `non-core' component may also contain a contribution from a minihalo if present.  We also note that in general the division between minihalo and jet/lobe may not be clear-cut if the minihalo is composed of electrons from the jet/lobe which have been redistributed by turbulence.

\begin{figure*}
    \centering
    \includegraphics[trim=3.3cm 0cm 2.6cm 0cm, clip, width=\columnwidth]{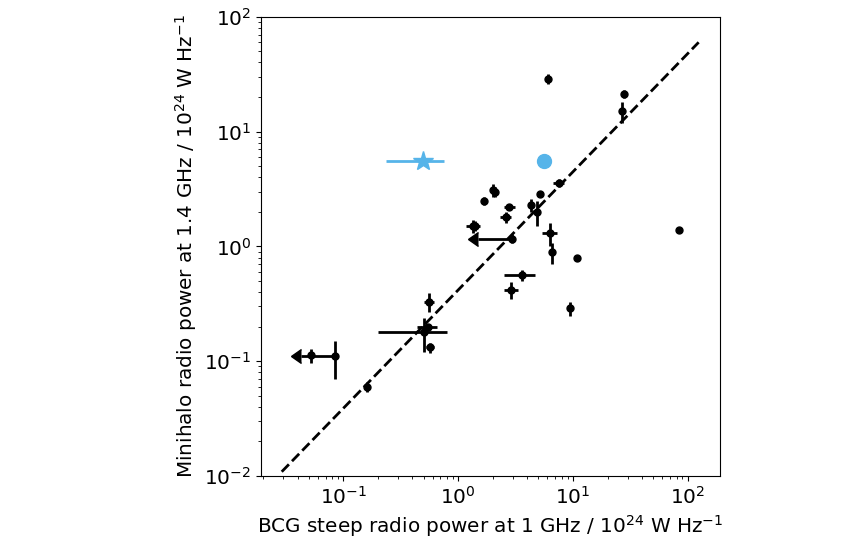}
    \includegraphics[trim=3.3cm 0cm 2.6cm 0cm, clip, width=\columnwidth]{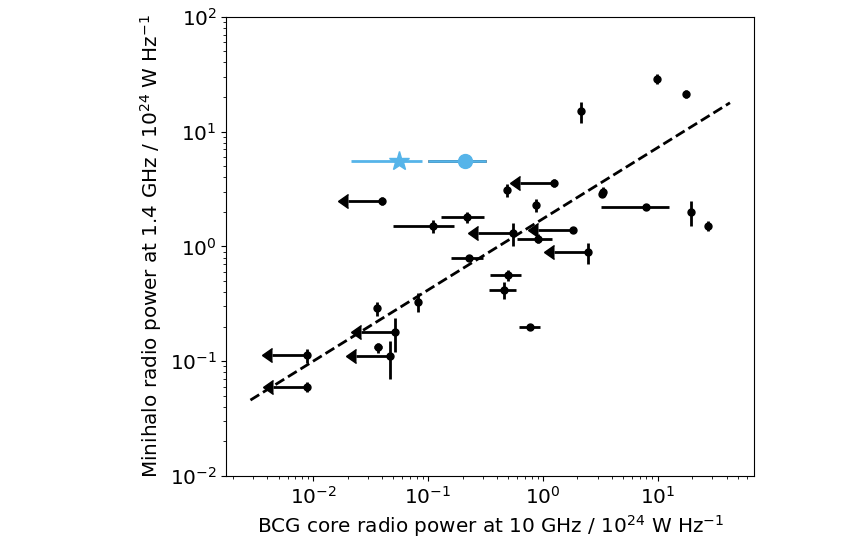}
    \caption{Updated correlations from \citet{2020MNRAS.499.2934R} between steep and core radio powers of the BCG and minihalo power.  The large blue circle in each plot shows the previous datapoint for RXJ1720.1+2638, relying on spectral separation of the components, and the blue star shows our updated datapoint using the spatially-separated estimates.}.
    \label{Fi:correlations}
\end{figure*}

With the current observation, we are now able to spatially decompose the RXJ1720.1+2638 BCG emission into core and jet component, as well as robustly exclude the minihalo component.  With these data in hand, we show the updated position of the BCG in the \citet{2020MNRAS.499.2934R} correlation plots in Fig.~\ref{Fi:correlations}, assuming their average spectral indices of $\alpha_{\mathrm{core}}=0.2$ and $\alpha_{\mathrm{steep}}=1.0$ to extrapolate our measurements to the appropriate frequency and calculate luminosity as we do not have spectral index measurements for the components given the narrow e-MERLIN frequency band.  The errorbars on the e-MERLIN measurements incorporate uncertainty on the spectral index of $\upDelta \alpha=0.2$ \citep{2015MNRAS.453.1201H} and assume a 60 per cent uncertainty on the core flux density due to its variability (they are dominated by the latter).  For simplicity here we use the slightly different cosmological parameters used by \citet{2020MNRAS.499.2934R}, but the difference in the calculated luminosities is not significant ($<2$\,per cent).

The updated datapoints for RXJ1720.1+2638 in both cases depart significantly from the correlation reported by \citet{2020MNRAS.499.2934R}.  This hints that the correlation may not be robust if other datapoints are similarly affected; we note that \citet{2022MNRAS.512.4210R} also found a similarly discrepant flux density measurement for MS\,1455.0+2232, another cool-core cluster with a minihalo.  With regard to the jet/lobe component, we also note that the argument that this represents an older population of electrons does not necessarily apply in this case since the jet is very close to the origin, implying that this emission is also young.

\section{Conclusions}

\begin{enumerate}
\item We have confirmed the presence of sub-kpc scale radio jets being produced by the AGN in the BCG of the cool-core cluster RXJ1720.1+2638.  
\item Based on the comparison with the integrated flux density measured at lower resolution, we estimate that we have recovered $\approx$\,60 per cent of the flux of the jets in our high-resolution observation and they may therefore extend out further than the measured length of $\approx$\,0.8\,kpc from the core.
\item Our spatially separated measurements may indicate that the significance of correlations between BCG radio luminosity and minihalo luminosity have been overestimated in previous work.
\item The presence of active jets may explain the relatively flat radio spectral index of the RXJ1720.1+2638 minihalo; further simulations are required to confirm this.
\end{enumerate}

\section*{Acknowledgements}

We thank an anonymous reviewer whose input helped improve the presentation of the paper.  e-MERLIN is a National Facility operated by the University of Manchester at Jodrell Bank Observatory on behalf of STFC.  YCP acknowledges support from a Rutherford Discovery Fellowship.

\section*{Data Availability}

Calibrated visibility data and/or images are available on reasonable request from the authors.



\bibliographystyle{mnras}
\bibliography{RXJ1720_BCG} 







\bsp	
\label{lastpage}
\end{document}